# Attribute Weighting with Adaptive NBTree for Reducing False Positives in Intrusion Detection

Dewan Md. Farid, and Jerome Darmont

ERIC Laboratory, University Lumière Lyon 2
Bat L - 5 av. Pierre Mendes, France
69676 BRON Cedex, France
dewanfarid@gmail.com, jerome.darmont@univ-lyon2.fr

Mohammad Zahidur Rahman

Department of Computer Science and Engineering
Jahangirnagar University
Dhaka – 1342, Bangladesh
rmzahid@juniv.edu

*Abstract*—In this paper, we introduce new learning algorithms for reducing false positives in intrusion detection. It is based on decision tree-based attribute weighting with adaptive naïve Bayesian tree, which not only reduce the false positives (FP) at acceptable level, but also scale up the detection rates (DR) for different types of network intrusions. Due to the tremendous growth of network-based services, intrusion detection has emerged as an important technique for network security. Recently data mining algorithms are applied on network-based traffic data and host-based program behaviors to detect intrusions or misuse patterns, but there exist some issues in current intrusion detection algorithms such as unbalanced detection rates, large numbers of false positives, and redundant attributes that will lead to the complexity of detection model and degradation of detection accuracy. The purpose of this study is to identify important input attributes for building an intrusion detection system (IDS) that is computationally efficient and effective. Experimental results performed using the KDD99 benchmark network intrusion detection dataset indicate that the proposed approach can significantly reduce the number and percentage of false positives and scale up the balance detection rates for different types of network intrusions.

## I.    INTRODUCTION

With the popularization of network-based services, intrusion detection systems (IDS) have become important tools for ensuring network security that is the violation of information security policy. IDS collect information from a variety of network sources using intrusion detection sensors, and analyze the information for signs of intrusions that attempt to compromise the confidentiality and integrity of networks [1]-[3]. Network-based intrusion detection systems (NIDS) monitor and analyze network traffics in the network for detecting intrusions from internal and external intruders [4]-[9]. Internal intruders are the inside users in the network with some authority, but try to gain extra ability to take action without legitimate authorization. External intruders are the outside users without any authorized access to the network that they attack. IDS notify network security administrator or automated intrusion prevention systems (IPS) about the network attacks, when an intruder try to break the network. Since the amount of audit data that an IDS needs to examine is very large even for a small network, several data mining algorithms, such as decision tree, naïve Bayesian classifier, neural network, Support Vector Machines, and fuzzy classification, etc [10]-[20] have been widely used by the IDS community for detecting known and unknown intrusions. Data mining based intrusion detection algorithms aim to solve the problems of analyzing the huge volumes of audit data and realizing performance optimization of detection rules [21]. But there are still some drawbacks in currently available commercial IDS, such as low detection accuracy, large number of false positives, unbalanced detection rates for different types of intrusions, long response time, and redundant input attributes.

A conventional intrusion detection database is complex, dynamic, and composed of many different attributes. The problem is that not all attributes in intrusion detection database may be needed to build efficient and effective IDS. In fact, the use of redundant attributes may interfere with the correct completion of mining task, because the information they added is contained in other attributes. The use of all attributes may simply increase the overall complexity of detection model, increase computational time, and decrease the detection accuracy of the intrusion detection algorithms. It has been tested that effective attributes selection improves the detection rates for different types of network intrusions in intrusion detection. In this paper, we present new learning algorithms for network intrusion detection using decision tree-based attribute weighting with adaptive naïve Bayesian tree. In naïve Bayesian tree (NBTree) nodes contain and split as regular decision-trees, but the leaves contain naïve Bayesian classifier. The proposed approach estimates the degree of attribute dependency by constructing decision tree, and considers the depth at which attributes are tested in the tree. The experimental results show that the proposed approach not only improves the balance detection for different types of network intrusions, but also significantly reduce the number and percentage of false positives in intrusion detection.

The rest of this paper is organized as follows. In Section II, we outline the intrusion detection models, architecture of data mining based IDS, and related works. In Section III, the basic concepts of feature selection and naïve Bayesian tree are introduced. In Section IV, we introduce the proposed algorithms. In Section V, we apply the proposed algorithms to the area of intrusion detection using KDD99 benchmark network intrusion detection dataset, and compare the results to other related algorithms. Finally, Section VI contains the conclusions with future works.

## II. INTRUSION DETECTION SYSTEM: IDS

### A. Misuse Vs. Anomaly Vs. Hybrid Detection Model

Intrusion detection techniques are broadly classified into three categories: misuse, anomaly, and hybrid detection model. Misuse or signature based IDS detect intrusions based on known intrusions or attacks stored in database. It performs pattern matching of incoming packets and/or command sequences to the signatures of known attacks. Known attacks can be detected reliably with a low false positive using misuse detection techniques. Also it begins protecting the computer/network immediately upon installation. But the major drawback of misuse-based detection is that it requires frequently signature updates to keep the signature database up-to-date and cannot detect previously unknown attacks. Misuse detection system use various techniques including rule-based expert systems, model-based reasoning systems, state transition analysis, genetic algorithms, fuzzy logic, and keystroke monitoring [22]-[25].

Anomaly based IDS detect deviations from normal behavior. It first creates a normal profile of system, network, or program activity, and then any activity that deviated from the normal profile is treated as a possible intrusion. Various data mining algorithms have been using for anomaly detection techniques including statistical analysis, sequence analysis, neural networks, artificial intelligence, machine learning, and artificial immune system [26]-[33]. Anomaly based IDS have the ability to detect new or previously unknown attacks, and insider attacks. But the major drawback of this system is large number of false positives. A false positive occurs when an IDS reports as an intrusion an event that is in fact legitimate network/system activity.

A hybrid or compound detection system detect intrusions by combining both misuse and anomaly detection techniques. Hybrid IDS makes decision using a "hybrid model" that is based on both the normal behavior of the system and the intrusive behavior of the intruders. Table I shows the comparisons of characteristics of misuse, anomaly, and hybrid detection models.

TABLE I. COMPARISONS OF INTRUSION DETECTION MODELS

| Characteristics | Misuse | Anomaly | Hybrid |
|---|---|---|---|
| Detection Accuracy | High (for known attacks) | Low | High |
| Detecting New Attacks | No | Yes | Yes |
| False Positives | Low | Very high | High |
| False Negatives | High | Low | Low |
| Timely Notifications | Fast | Slow | Rather Fast |
| Update Usage Patterns | Frequent | Not Frequent | Not Frequent |

### B. Architecture of Data Mining Based IDS

An IDS monitors network traffic in a computer network like a network sniffer and collects network logs. Then the collected network logs are analyzed for rule violations by using data mining algorithms. When any rule violation is detected, the IDS alert the network security administrator or automated intrusion prevention system (IPS). The generic architectural model of data mining based IDS is shown in Fig 1.

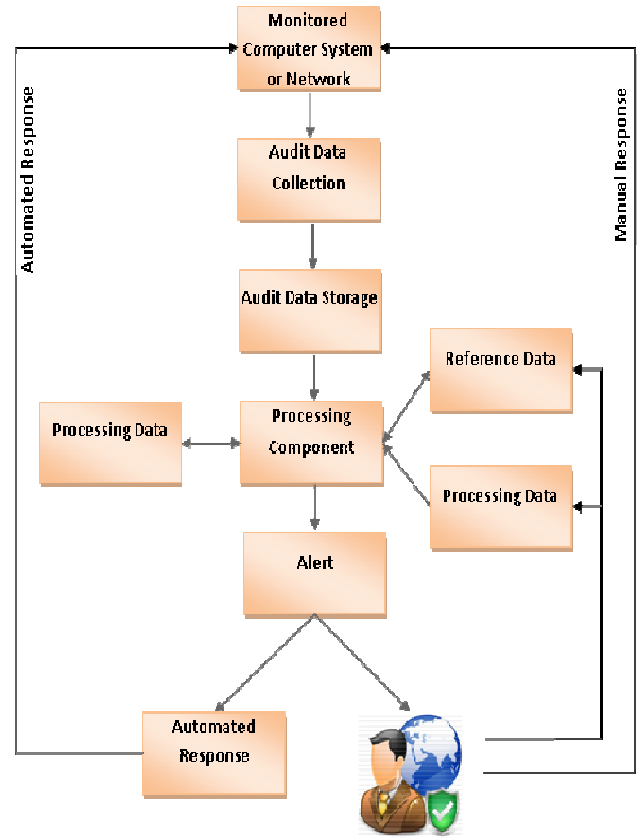

Figure 1. Organization of a generalized data mining based IDS

- Audit data collection: IDS collect audit data and analyzed them by the data mining algorithms to detect suspicious activities or intrusions. The source of the data can be host/network activity logs, command-based logs, and application-based logs.

- Audit data storage: IDS store the audit data for future reference. The volume of audit data is extremely large. Currently adaptive intrusion detection aims to solve the problems of analyzing the huge volumes of audit data and realizing performance optimization of detection rules.

- Processing component: The processing block is the heart of IDS. It is the data mining algorithms that apply for detecting suspicious activities. Algorithms for the analysis and detection of intrusions have been traditionally classified into two categories: misuse (or signature) detection, and anomaly detection.

- Reference data: The reference data stores information about known attacks or profiles of normal behaviors.

- Processing data: The processing element must frequently store intermediate results such as

information about partially fulfilled intrusion signatures.

- Alert: It is the output of IDS that notifies the network security officer or automated intrusion prevention system (IPS).

- System security officer or intrusion prevention system (IPS) carries out the prescriptions controlled by the IDS.

## C. Related Work

The concept of intrusion detection began with Anderson's seminal paper in 1980 [34] by introducing a threat classification model that develops a security monitoring surveillance system based on detecting anomalies in user behavior. In 1986, Dr. Denning proposed several models for commercial IDS development based on statistics, Markov chains, time-series, etc [35], [36]. In 2001, Lindqvist et al. proposed a rule-based expert system called eXpert-BSM for detecting misuse of host machine by analyzing activities inside the host in forms of audit trails [37], which generates detail reports and recommendations to the system administrators, and produces low false positives. Rules are conditional statements that derived from employing domain expert knowledge. In 2005, Fan et al. proposed a method to generate artificial anomalies into training dataset of IDS to handle both misuse and anomaly detection [38]. This method injects artificial anomaly data into the training data to help a baseline classifier distinguish between normal and anomalous data. In 2006, Bouzida et al. [39] introduced a supplementary condition to the baseline decision tree (DT) for anomaly intrusion detection. The idea is that instead of assigning a default class (normally based on probability distribution) to the test instance that is not covered by the tree, the instance is assigned to a new class. Then, instances with the new class are examined for unknown attack analysis. In 2009, Wu and Yen [21] applied DT and support vector machine (SVM) algorithm to built two classifiers for comparison by employing a sampling method of several different normal data ratios. More specifically, KDD99 dataset is split into several different proportions based on the normal class label for both training set and testing set. The overall evaluation of a classifier is based on the average value of results. It is reported that in general DT is superior to SVM classifier. In the same way, Peddabachigari et al. [40] applied DT and SVM for intrusion detection, and proven that decision tree is better than SVM in terms of overall accuracy. Particularly, DT much better in detecting user to root (U2R) and remote to local (R2L) network attacks, compared to SVM.

Naïve Bayesian (NB) classifier produces a surprising result of classification accuracy in comparison with other classifiers on KDD99 benchmark intrusion detection dataset. In 2001, Barbara et al. [41] proposed a method based on the technique called Pseudo-Bayes estimators to enhance the ability of ADAM intrusion detection system [42] in detecting new attacks and reducing false positives, which estimates the prior and posterior probabilities for new attacks by using information derived from normal instances and known attacks without requiring prior knowledge about new attacks. This study constructs a naïve Bayes Classifier to classify a given instance into a normal instance, known attack, or new attack. In 2004, Amor et al. [43] conducted an experimental study of the performance comparison between NB classifier and DT on KDD99 dataset. This experimental analysis reported that DT outperforms in classifying normal, denial of service (DoS), and R2L attacks, whereas NB classifier is superior in classifying Probe and U2R attacks. With respect to running time, the authors pointed out that NB classifier is 7 times faster than DT. Another naïve Bayes method for detecting signatures of specific attacks is motivated by Panda and Patra in 2007 [44]. From the experimental results implemented on KDD99 dataset, the authors give a conclusion that NB classifier performs back propagation neural network classifier in terms of detection rates and false positives. It is also reported that NB classifier produces a relatively high false positive. In a later work, the same authors Panda and Patra [45] in 2009, compares NB classifier with 5 other similar classifiers, i.e., JRip, Ridor, NNge, Decision Table, and Hybrid Decision Table, and experimental results shows that the NB classifier is better than other classifiers.

## III. FEATURE SELECTION AND ADAPTIVE NB TREE

### A. Feature Selection

Feature selection becomes indispensable for high performance intrusion detection using data mining algorithms, because irrelevant and redundant features may lead to complex intrusion detection model as well as poor detection accuracy. Feature selection is the process of finding a subset of features from total original features. The purpose of feature selection is to remove the irrelevant input features from the dataset for improving the classification accuracy. Feature selection in particularly useful in the application domains that introduce a large number of input dimensions like intrusion detection. Many data mining methods have been used for selecting important features from training dataset such as information gain based, gain ratio based, principal component analysis (PCA), genetic search, and classifier ensemble methods etc [46]-[53]. In 2009, Yang et al. [54] introduced a wrapper-based feature selection algorithm to find most important features from the training dataset by using random mutation hill climbing method, and then employs linear support vector machine (SVM) to evaluate the selected subset-features. Chen et al. [55] proposed a neural-tree based algorithm to identify important input features for classification, based on an evolutionary algorithm that the feature contributes more to the objective function will consider as an important feature.

In this paper, to select the important input attributes from training dataset, we construct a decision tree by applying ID3 algorithm in training dataset. The ID3 algorithm constructs decision tree using information theory [56], which choose splitting attributes from the training dataset with maximum information gain. Information gain is the amount of information associated with an attribute value that is related to the probability of occurrence. Entropy is the quantify information that is used to measure the amount of randomness from a dataset. When all data in a set belong to a single class, there is no uncertainty then the entropy is zero. The objective of ID3 algorithm is to iteratively partition the given dataset into sub-datasets, where all the instances in each final subset belong to the same class. The value for entropy is between 0 and 1 and reaches a maximum when the probabilities are all the same. Given probabilities $p_1, p_2,...,p_s$, where $\sum_{i=1} p_i = 1$;

$$Entropy: H(p_1,p_2,...p_s) = \sum_{i=1}^{s} (p_i \log(1/p_i)) \qquad (1)$$

Given a dataset, $D$, $H(D)$ finds the amount of sub-datasets of original dataset. When that sub-dataset is split into s new sub-datasets $S = \{D_1, D_2,...,D_s\}$, we can again look at the entropy of those sub-datasets. A subset is completely ordered if all instances in it are the same class. The ID3 algorithm calculates the gain by the equation "(2)".

$$Gain (D,S) = H(D) - \sum_{i=1}^{s} p(D_i)H(D_i) \qquad (2)$$

After constructing the decision tree from training dataset, we weight the attributes of training dataset by the minimum depth at which the attribute is tested in the decision tree. The depth of root node of the decision tree is 1. The weight for an attribute is set to $1/\sqrt{d}$, where $d$ is the minimum depth at which the attribute is tested in the tree. The weights of attributes that do not appear in the decision tree are assigned to zero.

### B. Naïve Bayesian Tree

Naïve Bayesian tree (NBTree) is a hybrid learning approach of decision tree and naïve Bayesian classifier. In NBTree nodes contain and split as regular decision-trees, but the leaves are replaced by naïve Bayesian classifier, the advantage of both decision tree and naïve Bayes can be utilized simultaneously [57]. Depending on the precise nature of the probability model, NB classifier can be trained very efficiently in a supervised learning. In many practical applications, parameter estimation for naïve Bayesian models uses the method of maximum likelihood. Suppose the training dataset, $D$ consists of predictive attributes $\{A_1, A_2,...,A_n\}$, where each attribute $A_i = \{A_{i1}, A_{i2},...,A_{ik}\}$ contains attribute values and a set of classes $C = \{C_1, C_2,...,C_n\}$. The objective is to classify an unseen example whose class value is unknown but values for attributes $A_1$ through $A_k$ are known. The aim of decision tree learning is to construct a tree model: $\{A_1, A_2,...,A_n\} \rightarrow C$. Correspondingly the Bayes theorem, if attribute $A_i$ is discrete or continuous, we will have:

$$P(C_j \mid A_{ij}) = \frac{P(A_{ij} \mid C_j)P(C_j)}{P(A_{ij})} \qquad (3)$$

Where $P(C_j|A_{ij})$ denote the probability. The aim of Bayesian classification is to decide and choose the class that maximizes the posteriori probability. Since $P(A_{ij})$ is a constant independent of $C$, then:

$$C^* = \underset{c \in C}{\arg\max} P(C_j \mid A_{ij})$$
$$= \underset{c \in C}{\arg\max} P(A_{ij} \mid C_j)P(C_j) \qquad (4)$$

Adaptive naïve Bayesian tree splits the dataset by applying entropy based algorithm and then used standard naïve Bayesian classifiers at the leaf node to handle attributes. It applies strategy to construct decision tree and replaces leaf node with naïve Bayesian classifier.

### IV. PROPOSED LEARNING ALGORITHM

#### A. Proposed Attribute Weighting Algorithm

In a given training data, $D = \{A_1, A_2,...,A_n\}$ of attributes, where each attribute $A_i = \{A_{i1}, A_{i2},...,A_{ik}\}$ contains attribute values and a set of classes $C = \{C_1, C_2,...,C_n\}$, where each class $C_j = \{C_{j1}, C_{j2},...,C_{jk}\}$ has some values. Each example in the training data contains weight, $w = \{w_1, w_2..., w_n\}$. Initially, all the weights of examples in training data have equal unit value that set to $w_i = 1/n$. Where $n$ is the total number of training examples. Estimates the prior probability $P(C_j)$ for each class by summing the weights that how often each class occurs in the training data. For each attribute, $A_i$, the number of occurrences of each attribute value $A_{ij}$ can be counted by summing the weights to determine $P(A_{ij})$. Similarly, the conditional probability $P(A_{ij} | C_j)$ can be estimated by summing the weights that how often each attribute value occurs in the class $C_j$ in the training data. The conditional probabilities $P(A_{ij} | C_j)$ are estimated for all values of attributes. The algorithm then uses the prior and conditional probabilities to update the weights. This is done by multiplying the probabilities of the different attribute values from the examples. Suppose the training example $e_i$ has independent attribute values $\{A_{i1}, A_{i2},...,A_{ip}\}$. We already know the prior probabilities $P(C_j)$ and conditional probabilities $P(A_{ik}|C_j)$, for each class $C_j$ and attribute $A_{ik}$. We then estimate $P(e_i | C_j)$ by

$$P(e_i \mid C_j) = P(C_j) \prod P(A_{ij} \mid C_j) \qquad (5)$$

To update the weight of training example $e_i$, we can estimate the likelihood of $e_i$ for each class. The probability that $e_i$ is in a class is the product of the conditional probabilities for each attribute value. The posterior probability $P(C_j | e_i)$ is then found for each class. Then the weight of the example is updated with the highest posterior probability for that example and also the class value is updated according to the highest posterior probability. Now, the algorithm calculates the information gain by using updated weights and builds a tree. After the tree construction, the algorithm initialized weights for each attributes in training data $D$. If the attribute in the training data is not tested in the tree then the weight of the attribute is initialized to 0, else calculates the minimum depth, $d$ that the attribute is tested at and initialized the weight of attribute to $1/\sqrt{d}$. Finally, the algorithm removes all the attributes with zero weight from the training data $D$. The main procedure of proposed algorithm is described as follows.

#### Algorithm 1: Attribute Weighting
Input: Training Dataset, $D$
Output: Decision tree, $T$
Procedure:
1. Initialize all the weights for each example in $D$, $w_i = 1/n$, where $n$ is the total number of the examples.

2. Calculate the prior probabilities $P(C_j)$ for each class $C_j$ in $D$. $P(C_j) = \dfrac{\sum\limits_{C_i} w_i}{\sum\limits_{i=1}^{n} w_i}$

3. Calculate the conditional probabilities $P(A_{ij} \mid C_j)$ for each attribute values in $D$. $P(A_{ij} \mid C_j) = \dfrac{P(A_{ij})}{\sum\limits_{C_i} w_i}$

4. Calculate the posterior probabilities for each example in $D$.
$$P(e_i \mid C_j) = P(C_j) \prod P(A_{ij} \mid C_j)$$

5. Update the weights of examples in $D$ with Maximum Likelihood (ML) of posterior probability $P(C_j|e_i)$;
$$w_i = P_{ML}(C_j|e_i)$$

6. Change the class value of examples associated with maximum posterior probability, $C_j = C_{i \rightarrow} P_{ML}(C_j|e_i)$.

7. Find the splitting attribute with highest information gain using the updated weights, $w_i$ in $D$.
Information Gain=
$$\left( -\sum_{j=1}^{k} \frac{\sum\limits_{i=C_i} w_i}{\sum\limits_{i=1}^{n} w_i} \log \frac{\sum\limits_{i=C_i} w_i}{\sum\limits_{i=1}^{n} w_i} \right) - \left( \sum_{i=1}^{n} \frac{\sum\limits_{i=C_i} w_i}{\sum\limits_{i=C_i} w_i} \log \left( \sum\limits_{i=C_{ij}} w_i \right) \right)$$

8. $T$ = Create the root node and label with splitting attribute.

9. For each branch of the $T$, $D$ = database created by applying splitting predicate to $D$, and continue steps 1 to 8 until each final subset belong to the same class or leaf node created.

10. When the decision tree construction is completed, for each attribute in the training data $D$: If the attribute is not tested in the tree then weight of the attribute is initialized to 0. Else, let $d$ be the minimum depth that the attribute is tested in the tree, and weight of the attribute is initialized to $1/\sqrt{d}$.

11. Remove all the attributes with zero weight from the training data $D$.

### B. Proposed Adaptive NBTree Algorithm

Given training data, $D$ where each attribute $A_i$ and each example $e_i$ have the weight value. Estimates the prior probability $P(C_j)$ and conditional probability $P(A_{ij} \mid C_j)$ from the given training dataset using weights of the examples. Then classify all the examples in the training dataset using these prior and conditional probabilities with incorporating attribute weights into the naïve Bayesian formula:

$$P(e_i \mid C_j) = P(C_j) \prod_{i=1}^{m} P(A_{ij} \mid C_j)^{W_i} \qquad (6)$$

Where $W_i$ is the weight of attribute $A_i$. If any example of training dataset is misclassified, then for each attribute $A_i$, evaluate the utility, $u(A_i)$, of a spilt on attribute $A_i$. Let $j = argmax_i(u_i)$, i.e., the attribute with the highest utility. If $u_j$ is not significantly better than the utility of the current node, create a NB classifier for the current node. Partition the training data $D$ according to the test on attribute $A_i$. If $A_i$ is continuous, a threshold split is used; if $A_i$ is discrete, a multi-way split is made for all possible values. For each child, call the algorithm recursively on the portion of $D$ that matches the test leading to the child. The main procedure of algorithm is described as follows.

**Algorithm 2: Adaptive NBTree**
Input: Training dataset $D$ of labeled examples.
Output: A hybrid decision tree with naïve Bayesian classifier at the leaves.
Procedure:

1. Calculate the prior probabilities $P(C_j)$ for each class $C_j$ in $D$. $P(C_j) = \dfrac{\sum\limits_{C_i} w_i}{\sum\limits_{i=1}^{n} w_i}$

2. Calculate the conditional probabilities $P(A_{ij} \mid C_j)$ for each attribute values in $D$. $P(A_{ij} \mid C_j) = \dfrac{P(A_{ij})}{\sum\limits_{C_i} w_i}$

3. Classify each example in $D$ with maximum posterior probability. $P(e_i \mid C_j) = P(C_j) \prod\limits_{i=1}^{m} P(A_{ij} \mid C_j)^{W_i}$

4. If any example in D is misclassified, then for each attribute $A_i$, evaluate the utility, $u(A_i)$, of a spilt on attribute $A_i$.

5. Let $j = argmax_i(u_i)$, i.e., the attribute with the highest utility.

6. If $u_j$ is not significantly better than the utility of the current node, create a naïve Bayesian classifier for the current node and return.

7. Partition the training data $D$ according to the test on attribute $A_i$. If $A_i$ is continuous, a threshold split is used; if $A_i$ is discrete, a multi-way split is made for all possible values.

8. For each child, call the algorithm recursively on the portion of $D$ that matches the test leading to the child.

## V. EXPERIMENTAL RESULTS AND ANALYSIS

### A. Dataset

Experiments have been carried out on KDD99 cup benchmark network intrusion detection dataset, a predictive model capable of distinguishing between intrusions and normal connections [58]. In 1998, DARPA intrusion detection evaluation program, a simulated environment was set up to acquire raw TCP/IP dump data for a local-area network (LAN) by the MIT Lincoln Lab to compare the performance of various intrusion detection methods. It was operated like a real environment, but being blasted with multiple intrusion attacks and received much attention in the research community of adaptive intrusion detection. The KDD99 dataset contest uses a version of DARPA98 dataset. In KDD99 dataset each example represents attribute values of a class in the network data flow, and each class is labeled either normal or attack. Examples in KDD99 dataset are represented with a 41 attributes and also labeled as belonging to one of five classes as follows: (1) Normal traffic; (2) DoS (denial of service); (3) Probe, surveillance and probing; (4) R2L, unauthorized access from a remote machine; (5) U2R, unauthorized access to local super user privileges by a local unprivileged user. In KDD99 dataset these four attack classes are divided into 22 different attack classes that tabulated in Table II.

TABLE II.  ATTACKS IN KDD99 DATASET

| 4 Main Attack Classes | 22 Attack Classes |
|---|---|
| Denial of Service (DoS) | back, land, neptune, pod, smurt, teardrop |
| Remote to User (R2L) | ftp_write, guess_passwd, imap, multihop, phf, spy, warezclient, warezmaster |
| User to Root (U2R) | buffer_overflow, perl, loadmodule, rootkit |
| Probing | ipsweep, nmap, portsweep, satan |

The input attributes in KDD99 dataset are either discrete or continuous values and divided into three groups. The first group of attributes is the basic features of network connection, which include the duration, prototype, service, number of bytes from source IP addresses or from destination IP addresses, and some flags in TCP connections. The second group of attributes in KDD99 is composed of the content features of network connections and the third group is composed of the statistical features that are computed either by a time window or a window of certain kind of connections. Table III shows the number of examples of 10% training data and 10% testing data in KDD99 dataset. There are some new attack examples in testing data, which is no present in the training data.

TABLE III.  NUMBER OF EXAMPLES IN TRAINING AND TESTING KDD99 DATA

| Attack Types | Training Examples | Testing Examples |
|---|---|---|
| Normal | 97277 | 60592 |
| Denial of Service | 391458 | 237594 |
| Remote to User | 1126 | 8606 |
| User to Root | 52 | 70 |
| Probing | 4107 | 4166 |
| Total Examples | 494020 | 311028 |

### B.  Performance Measures

In order to evaluate the performance of proposed learning algorithm, we performed 5-class classification using KDD99 network intrusion detection benchmark dataset and consider two major indicators of performance: detection rate (DR) and false positives (FP). DR is defined as the number of intrusion instances detected by the system divided by the total number of intrusion instances present in the dataset.

$$DR = \frac{Total\_\det ected\_attacks}{Total\_attacks} * 100 \qquad (7)$$

FP is defined as the total number of normal instances.

$$FP = \frac{Total\_misclassified\_process}{Total\_normal\_process} * 100 \qquad (8)$$

All experiments were performed using an Intel Core 2 Duo Processor 2.0 GHz processor (2 MB Cache, 800 MHz FSB) with 1 GB of RAM.

### C.  Experiment and analysis on Proposed Algorithm

Firstly, we use proposed algorithm 1 to perform attribute selection from training dataset of KDD99 dataset and then we use our proposed algorithm 2 for classifier construction. The performance of our proposed algorithm on 12 attributes in KDD99 dataset is listed in Table IV.

TABLE IV.  PERFORMANCE OF PROPOSED ALGORITHM ON KDD99 DATASET

| Classes | Detection Rates (%) | False Positives (%) |
|---|---|---|
| Normal | 100 | 0.04 |
| Probe | 99.93 | 0.37 |
| DoS | 100 | 0.03 |
| U2R | 99.38 | 0.11 |
| R2L | 99.53 | 6.75 |

Table V and Table VI depict the performance of naïve Bayesian (NB) classifier and C4.5 algorithm using the original 41 attributes of KDD99 dataset.

TABLE V.  PERFORMANCE OF NB CLASSIFIER ON KDD99 DATASET

| Classes | Detection Rates (%) | False Positives (%) |
|---|---|---|
| Normal | 99.27 | 0.08 |
| Probe | 99.11 | 0.45 |
| DoS | 99.68 | 0.05 |
| U2R | 64.00 | 0.14 |
| R2L | 99.11 | 8.12 |

TABLE VI.  PERFORMANCE OF C4.5 ALGORITHM USING KDD99 DATASET

| Classes | Detection Rates (%) | False Positives (%) |
|---|---|---|
| Normal | 98.73 | 0.10 |
| Probe | 97.85 | 0.55 |
| DoS | 97.51 | 0.07 |
| U2R | 49.21 | 0.14 |
| R2L | 91.65 | 11.03 |

Table VII and Table VIII depict the performance of NB classifier and C4.5 using reduces 12 attributes.

TABLE VII.  PERFORMANCE OF NB CLASSIFIER USING KDD99 DATASET

| Classes | Detection Rates (%) | False Positives (%) |
|---|---|---|
| Normal | 99.65 | 0.06 |
| Probe | 99.35 | 0.49 |
| DoS | 99.71 | 0.04 |
| U2R | 64.84 | 0.12 |
| R2L | 99.15 | 7.85 |

TABLE VIII.  PERFORMANCE OF C4.5 ALGORITHM USING KDD99 DATASET

| Classes | Detection Rates (%) | False Positives (%) |
|---|---|---|
| Normal | 98.81 | 0.08 |
| Probe | 98.22 | 0.51 |
| DoS | 97.63 | 0.05 |
| U2R | 56.11 | 0.12 |
| R2L | 91.79 | 8.34 |

We also compare the intrusion detection performance among Support Vector Machines (SVM), Neural Network (NN), Genetic Algorithm (GA), and proposed algorithm on KDD99 dataset that tabulated in Table IX [59], [60].

TABLE IX.  COMPARISON OF SEVERAL ALGORITHMS

| | SVM | NN | GA | Proposed Algorithm |
|---|---|---|---|---|
| Normal | 99.4 | 99.6 | 99.3 | 99.93 |
| Probe | 89.2 | 92.7 | 98.46 | 99.84 |
| DoS | 94.7 | 97.5 | 99.57 | 99.91 |
| U2R | 71.4 | 48 | 99.22 | 99.47 |
| R2L | 87.2 | 98 | 98.54 | 99.63 |

## VI. CONCLUSIONS AND FUTURE WORKS

This paper presents a hybrid approach to intrusion detection based on decision tree-based attribute weighting with naïve Bayesian tree, which is suitable for analyzing large number of network logs. The main propose of this paper is to improve the performance of naïve Bayesian classifier for network intrusion detection systems (NIDS). The experimental results manifest that proposed approach can achieve high accuracy in both detection rates and false positives, as well as balanced detection performance on all four types of network intrusions in KDD99 dataset. The future works focus on applying the domain knowledge of security to improve the detection rates for current attacks in real time computer network, and ensemble with other mining algorithms for improving the detection rates in intrusion detection.

### ACKNOWLEDGMENT

Support for this research received from ERIC Laboratory, University Lumière Lyon 2 – France, and Department of Computer Science and Engineering, Jahangirnagar University, Bangladesh.

## AUTHORS PROFILE

**Dewan Md. Farid** was born in Dhaka, Bangladesh in 1979. He is currently a research fellow at ERIC Laboratory, University Lumière Lyon 2 - France. He obtained B.Sc. Engineering in Computer Science and Engineering from Asian University of Bangladesh in 2003 and Master of Science in Computer Science and Engineering from United International University, Bangladesh in 2004. He is pursuing Ph.D. in the Department of Computer Science and Engineering, Jahangirnagar University, Bangladesh. He is a faculty member in the Department of Computer Science and Engineering, United International University, Bangladesh. He is a member of IEEE and IEEE Computer Society. He has published 10 international research papers including two journals in the field of data mining, machine learning, and intrusion detection.

**Jérôme Darmont** received his Ph.D. in computer science from the University of Clermont-Ferrand II, France in 1999. He joined the University of Lyon 2, France in 1999 as an associate professor, and became full professor in 2008. He was head of the Decision Support Databases research group within the ERIC laboratory from 2000 to 2008, and has been director of the Computer Science and Statistics Department of the School of Economics and Management since 2003. His current research interests mainly relate to handling so-called complex data in data warehouses (XML warehousing, performance optimization, auto-administration, benchmarking...), but also include data quality and security as well as medical or health-related applications.

**Mohammad Zahidur Rahma** is currently a Professor at Department of Computer Science and Engineering, Jahangirnagar University, Banglasesh. He obtained his B.Sc. Engineering in Electrical and Electronics from Bangladesh University of Engineering and Technology in 1986 and his M.Sc. Engineering in Computer Science and Engineering from the same institute in 1989. He obtained his Ph.D. degree in Computer Science and Information Technology from University of Malaya in 2001. He is a co-author of a book on E-commerce published from Malaysia. His current research includes the development of a secure distributed computing environment and e-commerce.